\documentclass[apjl]{emulateapj}                                                 
\usepackage{color}  
\usepackage{rotating} 

\shorttitle{A fourth  21-cm absorber towards MG\,J0414+0534}
\shortauthors{A. Tanna et al.}

\def\kms{km ${\rm s}^{-1}$}

\def\ch2{$\chi^2$}

\def\kms {\hbox{${\rm km\ s}^{-1}$}}

\def\scm  {$\hbox{{\rm cm}}^{-2}$}    

\def \AL {$\alpha $}     
\def \HI {H{\sc \,i}}

\def\lapp{\ifmmode\stackrel{<}{_{\sim}}\else$\stackrel{<}{_{\sim}}$\fi}
\def\gapp{\ifmmode\stackrel{>}{_{\sim}}\else$\stackrel{>}{_{\sim}}$\fi}

\begin{document}


\title{A fourth \HI\ 21-cm absorption system in the sight-line of MG\,J0414+0534: a record for intervening  absorbers}


\author{A. Tanna\altaffilmark{1}, S. J. Curran\altaffilmark{2,3}, M. T. Whiting\altaffilmark{4},  J. K.  Webb\altaffilmark{1} and C. Bignell\altaffilmark{5}}
\affil{School of Physics, University of New South Wales, Sydney NSW 2052, Australia}
\affil{Sydney Institute for Astronomy, School of Physics, The University of Sydney, NSW 2006, Australia}
\email{sjc@physics.usyd.edu.au}
\affil{ARC Centre of Excellence for All-sky Astrophysics (CAASTRO)}
\affil{CSIRO Astronomy and Space Science, PO Box 76, Epping NSW 1710, Australia}
\affil{National Radio Astronomy Observatory, P.O. Box 2, Rt. 28/92 Green Bank, WV 24944-0002, USA}

\begin{abstract}
  We report the detection of a strong \HI\ 21-cm absorption system at $z=0.5344$, as well as a candidate system at $z=0.3389$, in the sight-line towards
  the $z=2.64$ quasar MG\,J0414+0534. This, in addition to the absorption at the host redshift and the other two
  intervening absorbers, takes the total to four (possibly five). The previous maximum number of 21-cm absorbers detected along a single
  sight-line is two and so we suspect that this number of gas-rich absorbers is in some way related to the very red colour of the
  background source. Despite this, no molecular gas (through OH absorption) has yet been detected at any of the 21-cm
  redshifts, although, from the population of 21-cm absorbers as a whole, there is evidence for a weak correlation between the atomic line strength and the
  optical--near-infrared colour. In either case, the fact that so many gas-rich galaxies (likely to be damped Lyman-$\alpha$ absorption systems) have been found along a single
  sight-line towards a highly obscured source may have far reaching implications for the population of faint galaxies not
  detected in optical surveys, a possibility which could be addressed through future wide-field absorption line surveys with the Square Kilometre Array.
\end{abstract}


\section{Introduction}
\label{intro}
Absorption of the 21-cm electron spin-flip transition of neutral hydrogen (\HI) is a powerful probe of the reservoir of
star-forming material in the early Universe. At high redshift, 21-cm absorption can provide important insight into star
formation rates and galaxy evolution at a time when chemical abundances were markedly different to the present
day. Furthermore, in combination with other absorbing species, \HI\ 21-cm can provide measurements of the fundamental
constants at large look-back times (\citealt{cdk04} and references therein). In five cases, OH 18-cm absorption has been found
coincident with the \HI\ \citep{cdn99,kc02a,kcdn03,kcl+05}\footnote{For all of these the full-width half maximum (FWHM) of 
the OH profile is similar to that of the \HI\ \citep{cdbw07}.} and the hydroxyl radical is of particular interest for
measurement of the constants since it allows highly sensitive measurements from a single absorbing species \citep{dar03}.

However, the detection of either transition at $z > 0.1$ is a rare occurrence. For \HI\ 21-cm, 80 systems have been
detected\footnote{Half of which occur in systems intervening the sight-lines to more distant radio sources, with the
  other half associated with a source's host galaxy (see \citealt{cur09a,cw10} and references therein).} and, despite
much searching (\citealt{cwm+10} and references therein), the detection of OH 18-cm is rarer yet with only the
aforementioned five detections to date.\footnote{Three of which are intervening and two associated absorbers.}
\citet{cwm+06,cwc+11} have shown that the molecular fraction in the known molecular absorbers is correlated with the optical--near-infrared
colour, thus indicating that their colours are due to the presence of dust required to shield the molecular gas from
the ambient UV field. Therefore, selecting objects for which an optical redshift is available selects against those with
a high molecular fraction, making the detection of  molecular absorbers difficult.

Hence, in an attempt to increase the number of redshifted detections of both species, we have performed full spectral
scans with the Green Bank Telescope (GBT) toward five highly reddened (optical--near-infrared colours of $V - K > 6$)
radio-loud objects (Tanna et al., in prep.). For the reddest ($V - K = 10.26$) of the targets, the $z = 2.64$ quasar
MG\,J0414+0534 (4C\,+05.19), three 21-cm absorption systems have already been detected (see Table~\ref{abs}). In this
letter we report the detection of a further absorber at a redshift of $z = 0.534$, as well as candidate system at $z = 0.339$.
  
\section{The detection of one (and possibly two) new absorption feature(s)}
\label{oadr}

\begin{figure}                        
\centering \includegraphics[trim=0 0 0 0, angle=0,scale=0.46]{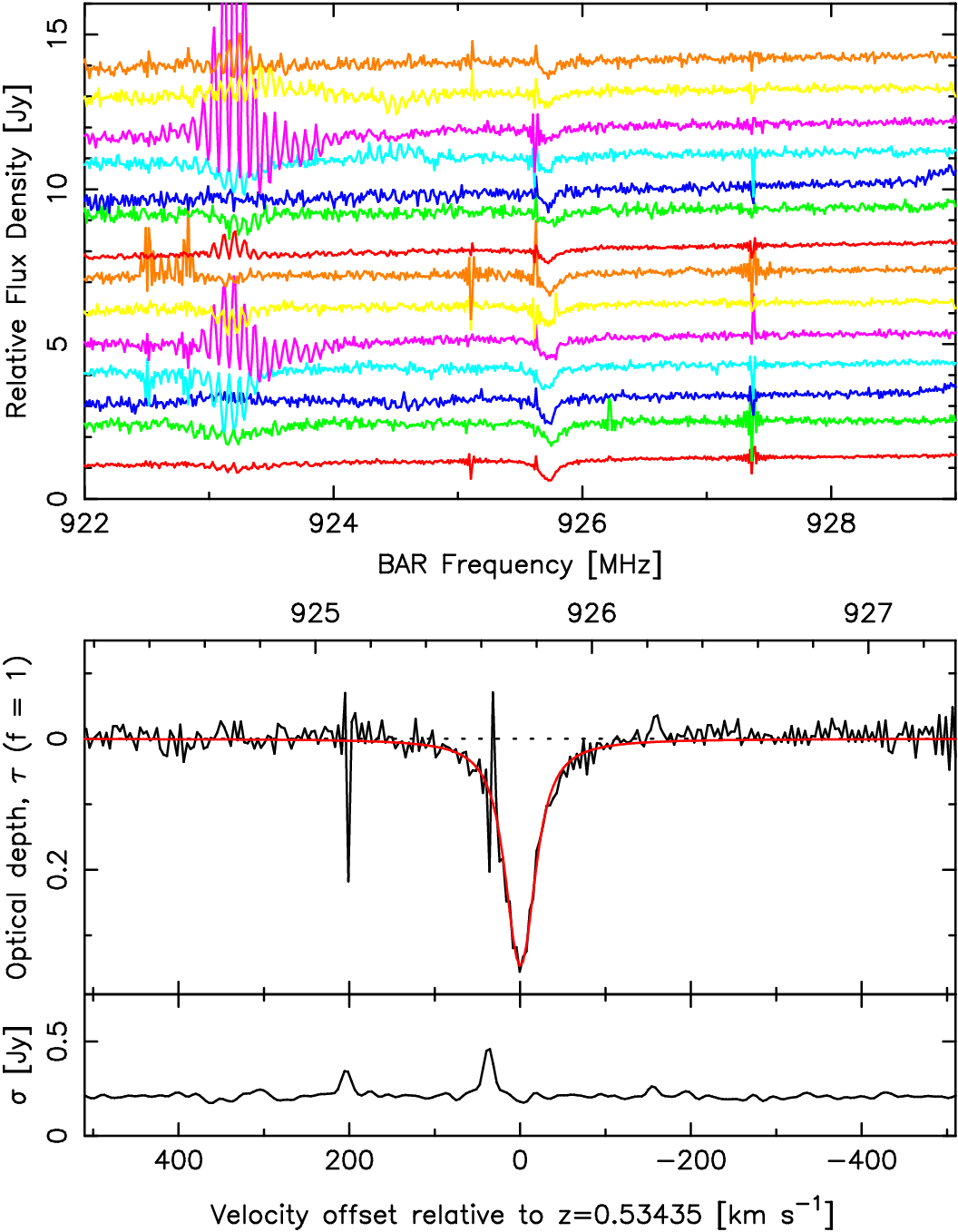}
\caption{The absorption feature near 926 MHz toward J0414+0534. Top: As seen in individual 146.3 sec exposure
scans of each linear polarisation. 
Bottom:  Detail showing the averaged spectrum. In this and Fig. \ref{5th-stacked}, the spectrum is shown at
  the observed spectral resolution of 3.4 \kms\ and a first order polynomial removed from the bandpass, with the
  best fit Voigt profile overlain. The lower
  panel shows the standard deviation of each channel from the mean across all individual scans.
}
\label{4th-stacked}
\end{figure}

We have now completed the analysis of the data along the entire redshift space towards MG\,J0414+0534 (see \citealt{cwt+11} for details), upon which we 
find a further two possible absorption profiles, near 926 MHz and 1061 MHz. 
The absorption feature close to 926 MHz was the least subject to radio frequency interference (RFI), which appears as narrow lines and amplitude fluctuations. 
These can be seen in Fig.~\ref{4th-stacked} (top), which essentially constitutes a low-resolution time-lapse series of the individual 146.3 sec exposure scans.

In the final averaged spectrum (Fig.~\ref{4th-stacked}, bottom), a persistent absorption feature was apparent, which
maintains consistency throughout the observations, with GBT RFI monitoring indicating that this part of the band is
predominantly clear of interference.  There remains, however, a two channel wide spike redshifted by $\approx40$ \kms\
from the peak of the profile. The RFI nature of this is confirmed in spectral animations of contiguous 5 sec
integrations, where like the other RFI features in the band, the spike is seen to fluctuate between its minimum and maximum amplitudes
on time-scales of a minute. For the sake of transparency, we do not flag the spike out of final averaged spectrum,
although we blank and interpolate over the affected channels for the measurements described in Sect.~\ref{hi4}.
Given the profile
shape\footnote{For the other possible candidate, OH 18-cm, we expect two features separated by $1.9572/(z+1)$ MHz.} and
its strength, we believe that the feature must arise from 21-cm absorption at $z=0.534$.

Close to 1061 MHz, the dominant RFI was apparent as ``packets'', one of which impinges on one side of a possible absorption
feature. RFI monitoring indicates that these packets (spaced $\sim$1 MHz apart) are due to aircraft radar and much of
the data are heavily affected. 
\begin{figure}
\centering \includegraphics[trim=0 0 0 0, angle=0,scale=0.46]{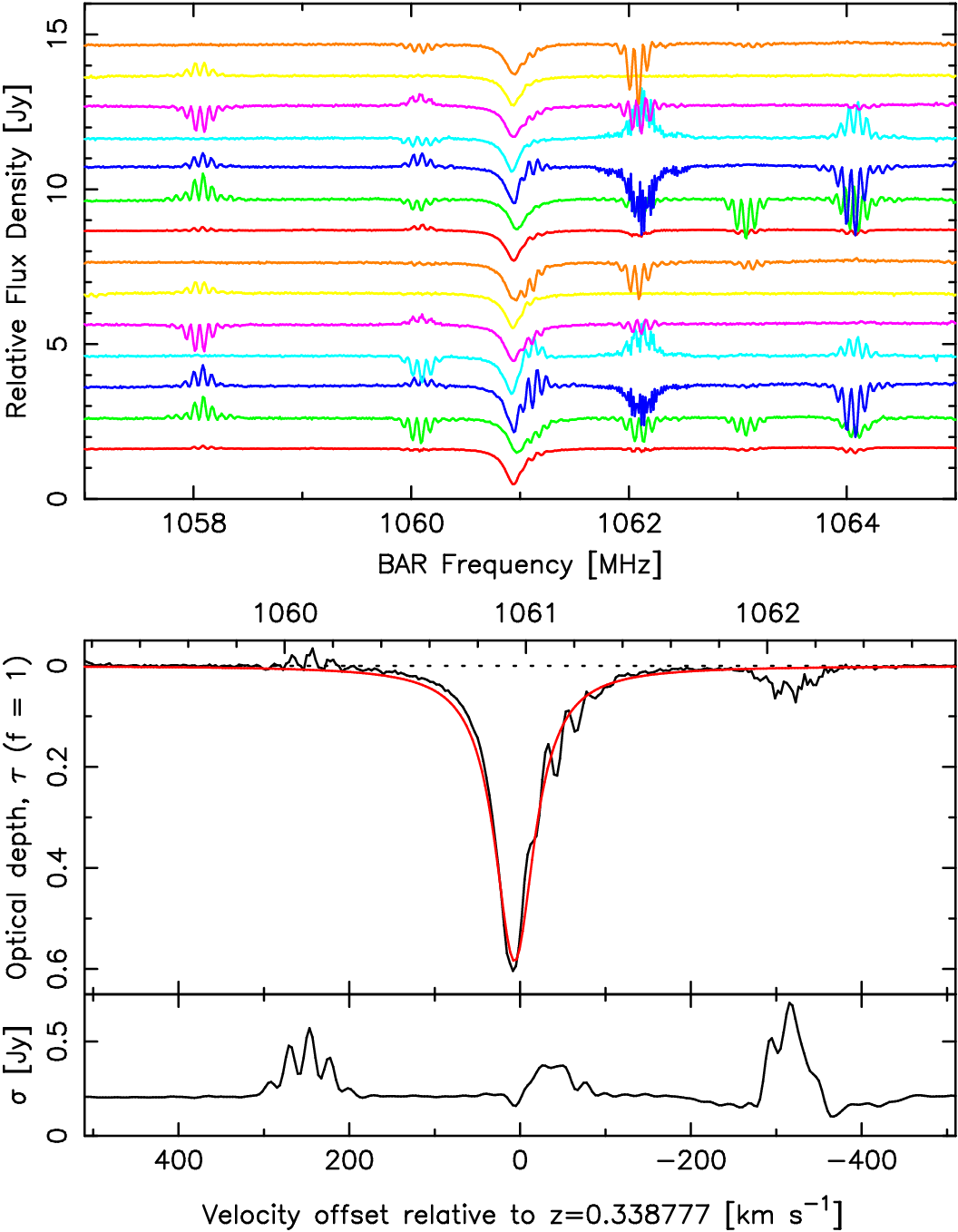}
\caption{As per Fig. \ref{4th-stacked}, but for the feature near 1061 MHz.}
\label{5th-stacked}
\end{figure}
Again, through the animation of contiguous 5 sec integrations, the packets behave in a similar fashion to the RFI spikes
close to 925 MHz, displaying time dependent fluctuations.  Flagging the data most strongly affected leaves
approximately one third of the scans, allowing us to retain some data at all
frequencies, particularly on the blue-shifted side of the line.  While the RFI packets are still apparent in the final
averaged spectrum (Fig. \ref{5th-stacked}, bottom), it is clear that the absorption feature at 1061 MHz maintains a
stable, deep profile shape over the course of the observations (even beneath the RFI). Furthermore, the packets vary
regularly in time, flipping between positive, negative and zero fluxes, whereas the absorption profile is consistent and
has an identical appearance in two different IFs (0.9--1.1 GHz and 1.05--1.25 GHz).
 The stability of the putative line is confirmed in the standard deviation spectrum, which identifies time
varying features, where the peak is not coincident with the most variable channels (Fig. \ref{5th-stacked}, bottom).
However, given that the near-by RFI is not restricted to one or two channels, this feature requires confirmation from an
independent observation.

\section{Results and discussion}
\label{randd}
\subsection{Absorption by neutral hydrogen}
\label{hi4}

The mean frequency of the first feature (Fig. \ref{4th-stacked}) is $925.738\pm0.009$ MHz, which for \HI\ 21-cm gives,
from the flagged data, a flux averaged mean redshift of $z = 0.53435\pm0.00002$, cf. the peak redshift of $z_{\rm peak}
= 0.53437\pm0.00002$, obtained from the deepest channel in the peak of the profile.  There is a decrease in the
bandpass response below 950 MHz, giving a continuum flux of $S = 1.82\pm0.05$ Jy, cf. the 2.3 Jy seen over the rest of the band,
which yields an observed peak optical depth of $\tau_{\rm obs}\equiv\Delta S/S = 0.30\pm0.04$. Unlike the
vast majority of redshifted 21-cm absorption, this peak depth means that the optically thin approximation cannot be
applied (where $\Delta S \lapp0.3\,S$), i.e. $\tau\equiv-\ln\left(1-{\tau_{\rm obs}}/{f_{\rm HI}}\right)\approx
{\tau_{\rm obs}}/{f_{\rm HI}}$. Since, by definition, the covering factor $f_{\rm HI}\leq1$, the peak optical depth is
then $\tau_{_{\rm peak}} \geq 0.36\pm0.06$ and the velocity integrated optical depth $\int\!\tau dv
\geq16\pm 3$ \kms.
This gives a column density of $N_{\rm HI} \geq 2.8\times10^{19}.\,T_{\rm spin}$, where $T_{\rm spin}$ [K] is the mean harmonic spin
temperature of the gas.
 
The mean frequency of the second  feature (Fig. \ref{5th-stacked}) is $1060.975\pm0.004$ MHz, which for \HI\ 21-cm  gives a redshift of 
$z = 0.338777\pm0.000006$, cf. the peak redshift of $z_{\rm peak} = 0.33882\pm0.00002$.
The peak depth of the line is $\Delta S = 1.059$ Jy and the continuum flux is $S = 2.331\pm0.015$ Jy, giving a peak $\tau_{\rm obs} = 0.45\pm0.01$.
Again, the optically thin approximation is not applicable and so $\tau_{\rm peak} \geq 0.61\pm0.02$, giving
$\int\!\tau dv \geq 37\pm 1$ \kms\ and $N_{\rm HI} \geq 6.7\times10^{19}.\,T_{\rm spin}$.
If genuine, this would be the second deepest redshifted 21-cm feature yet found\footnote{The deepest being $\tau_{_{\rm peak}} \geq0.71$ at $z
= 0.524$ towards AO\,0235+164 \citep{rbb+76}.}, although given our concerns about its authenticity (Sect.~\ref{oadr}) this requires confirmation.

We note there are apparent similarities in the shapes of the two absorption profiles as a function of velocity, although,
in addition to different depths, the widths exhibit a slight difference, with the 926 MHz profile having a full-width
half maximum of FWHM$\,\approx44$ \kms, cf. FWHM\,$\approx61$ \kms\ at 1061 MHz. Since the profile shapes are not consistent in
frequency space, we believe it unlikely that these are due to an instrumental artifact, which lends some weight to the
reality of the 1061 MHz feature. The shape of each line has much broader wings than those produced by a single Gaussian
profile and are best fit by a single Voigt profile, 
which we interpret as the effect of pressure broadening convolved with
 the velocity dispersion of the gas.

Lastly, in Table \ref{abs} we list the derived column densities together with the other  21-cm absorbers thus far found towards J0414+0534.
\begin{table*}
\centering
\begin{minipage}{150mm}  
   \caption{The absorbing systems thus far found along the sight-line to J0414+0534. The first column lists the redshift, followed by the feature
with which the absorption is associated. $N_{\rm HI}$ and $N_{\rm OH}$ give the \HI\ 21-cm and OH 18-cm line strengths, respectively,
followed by the normalised OH line strength, in terms of $(f_{\rm HI}/f_{\rm OH}).(T_{\rm ex}/T_{\rm spin})$. 
The last column gives the reference for the discovery of the absorption and the line strengths. 
\label{abs}}
\begin{tabular}{@{}l l c c  c l @{}}
  \hline
Redshift & Location &   $N_{\rm HI}$ [\scm] & $N_{\rm OH}$ [\scm]   & $N_{\rm OH}/N_{\rm HI}$ & Reference\\
 \hline
2.63647 & Host galaxy &  $7.5\times10^{18}.\,(T_{\rm spin}/f_{\rm HI})$ & $\lapp3\times10^{14}\,(T_{\rm ex}/f_{\rm OH})^{\dagger}$& $\lapp2\times10^{-4}$ & \citet{mcm98} \\
0.95974 & Lensing galaxy &$1.6\times10^{18}.\,(T_{\rm spin}/f_{\rm HI})$ &  $\leq4.9\times10^{12}.\,(T_{\rm ex}/f_{\rm OH})$&  $\leq9\times10^{-6}$& \citet{cdbw07} \\ 
0.53435 &     ---                      &  $\geq2.8\times10^{19}.\,T_{\rm spin}$  & $\leq1.3\times10^{14}.\,(T_{\rm ex}/f_{\rm OH})$&  $\leq5\times10^{-6}$& This paper \\
0.37895 & Object X & $2.9\times10^{19}.\,(T_{\rm spin}/f_{\rm HI})$ &  $\leq3.3\times10^{13}.\,(T_{\rm ex}/f_{\rm OH})$&  $\leq2\times10^{-6}$& \citet{cwt+11} \\ 
0.33878 &     ---         &$\geq6.7\times10^{19}.\,T_{\rm spin}$ &  $\leq5.0\times10^{14}.\,(T_{\rm ex}/f_{\rm OH})$&  $\leq8\times10^{-6}$& This paper$^{\ast}$ \\
\hline
\end{tabular}
{Notes: $^{\dagger}$OH limit from \citet{cwt+11}, $^{\ast}$if genuine 21-cm absorption (Sect. \ref{oadr}).}
\end{minipage}
\end{table*} 
Even the weakest (that arising in the lensing galaxy) is likely to qualify as a damped Lyman-$\alpha$ absorption system (DLA), requiring only
$(T_{\rm spin}/f_{\rm HI}) \geq 125$ K to reach the defining $N_{\rm HI}\geq2\times10^{20}$ \scm. That is, all of the absorbers so far
found along the sight-line towards J0414+0534 are likely to be associated with gas-rich galaxies.

\subsection{Optical counterparts}

Confirmation of 21-cm detections can be done through optical imaging and spectroscopy, either by finding absorption
lines in the quasar spectrum at matching redshifts or by identifying the absorbing galaxy.  Detecting common optical
absorption lines, such as Mg\,\textsc{ii}, from these systems 
is made difficult by the extremely red colour of the quasar's optical spectrum: For $z = 0.5343$ and $z=0.3388$ the
Mg\,\textsc{ii} doublet would appear at 429~nm and 375~nm respectively, where the quasar flux is very low
\citep{htl+92,lejt95}.  Furthermore, \citet{mcm98} detect strong 21-cm absorption in the host galaxy and any associated
Lyman-$\alpha$ absorption (centred on 442.6~nm at $z=2.64$) would likely conceal the near-by $z=0.5343$ Mg\,\textsc{ii}
line. 
 
There is evidence from the spectrum of \citet{htl+92} of the $\lambda\lambda5891,5897$ Na\,\textsc{i} doublet at
$z=0.3388$. This identification is made difficult by both the low-resolution of the published spectrum and the presence
of the nearby Ca\,\textsc{ii} H+K line from the lensing galaxy at $z=0.958$. It is, however, suggestive of a galaxy at
the redshift of the 1061 MHz putative detection. The same Na\,\textsc{i} line at $z=0.534$ coincides with strong
sky-line subtraction residuals in the \citeauthor{htl+92} spectrum.

Identifying the corresponding absorbing galaxy is also difficult. No published spectroscopy exists of neighbouring field
galaxies, although \citet{tk99} presented long-slit spectroscopy of the lensing galaxy, taken with the slit placed
across "Object X", the closest field galaxy. \citet{cwt+11} identified two peaks in the low-resolution spectrum at the
wavelengths expected for [O\,\textsc{iii}] at z=0.3789. It is likely, then, that this object does not correspond to
either of the detections presented here. HST images indicate numerous nearby galaxies, but none have redshift
measurements or estimates.\footnote{There are two galaxies, u2fl1\#038 and u2fl1\#044, at relatively large separations
  ($1.3-1.5$ arcmin), although these are extremely red objects  and so likely to be at $z>1$ \citep{yt03}.} Detailed, deep
multi-object and/or integral-field spectroscopy of this field would be required to fully resolve the identification of
the specific absorbing galaxies.

 \subsection{The incidence of multiple intervening absorbers along a single sight-line}
\label{sec:inc}

 The detection of at least one new absorber now gives at least three intervening systems along the sight-line to
 J0414+0534 (Table \ref{abs}). All of these systems are likely to be DLAs (Sect. \ref{hi4}), indicating a similar
 sight-line to the $z=3.02$ quasar CTQ\,247, towards which \citet{lmmm01} report the detection of four DLAs.  However,
 at $z=2.55$, $2.59$ and $2.62$, three of these arise in a single broad feature and the fourth, at $z=1.91$, is inferred
 from the metal lines. With a rest-frame equivalent width of $W_{\rm r}^{\lambda1216} = 6.0$~\AA\ for the Lyman-\AL\
 line, this is more likely to be a Lyman-limit system than a DLA. 

 In order to quantify how rare an occurrence the presence of at lease three DLAs  along a single sight-line 
 is, we use the \citet{ppb06} sample of strong (${\rm W}_{\rm r}^{\lambda2796} > 1$~\AA) Mg{\sc \,ii}
 absorbers. These are obtained from SDSS DR3 observations of 45\,023 QSO sight-lines, which span $0.35
 < z_{\rm MgII} < 2.3$, a similar redshift range to the 21-cm observations towards J0414+0534 at $z=2.64$.  However, the
 GBT scan below 700 MHz is completely dominated by RFI and so not sensitive to any absorption between the lensing
 galaxy and J0414+0534 (Tanna et al., in prep.), restricting the redshifts scanned to $z\lapp1$.  The total number of sight-lines probed for a given
 redshift value is in excess of 15\,000 for much of this range (see Fig. 2 of \citealt{ppb06}), peaking at 22\,000 --
 23\,000 for $0.6<z<0.9$.  Of these, there are 2564 unique sight-lines which exhibit ${\rm W}_{\rm r}^{\lambda2796} >
 1$~\AA\ absorption at $z\leq1$. Only 78 of these contain two distinct\footnote{Separated by at least 10\,000 \kms (or
   $\Delta z\gapp0.03$).} absorbers, with a further four sight-lines containing three absorbers.
 
Restricting this to DLA strength absorbers  (${\rm W}_{\rm r}^{\lambda2796} \gapp 3$~\AA, assuming the ${\rm W}_{\rm
 r}^{\lambda2796}$--$N_{\rm HI}$ relation of \citealt{mc09}, see also \citealt{ctp+07}), there are 
81 sight-lines which contain a DLA, but none with more than one. 
Zero sight-lines with at least two DLA strength absorbers out of $\gapp15\,000$ sight-lines means that,
with at least three distinct DLAs at $z<1$,  the sight-line towards J0414+0534 is unprecedented.
 Our use of SDSS QSOs \citep{ppb06} selects against reddened sight-lines (see Fig. 1 of \citealt{cwc+11}),
where dust extinction is low, and so extinction may be significant in the sight-line of J0414+0534, given
the very red colour and the high incidence of cold, neutral absorbing gas.

\subsection{Reddening of the quasar light}
\subsubsection{Reddening by dust associated with molecular gas}

\begin{figure*}
\centering \includegraphics[angle=270,scale=0.65]{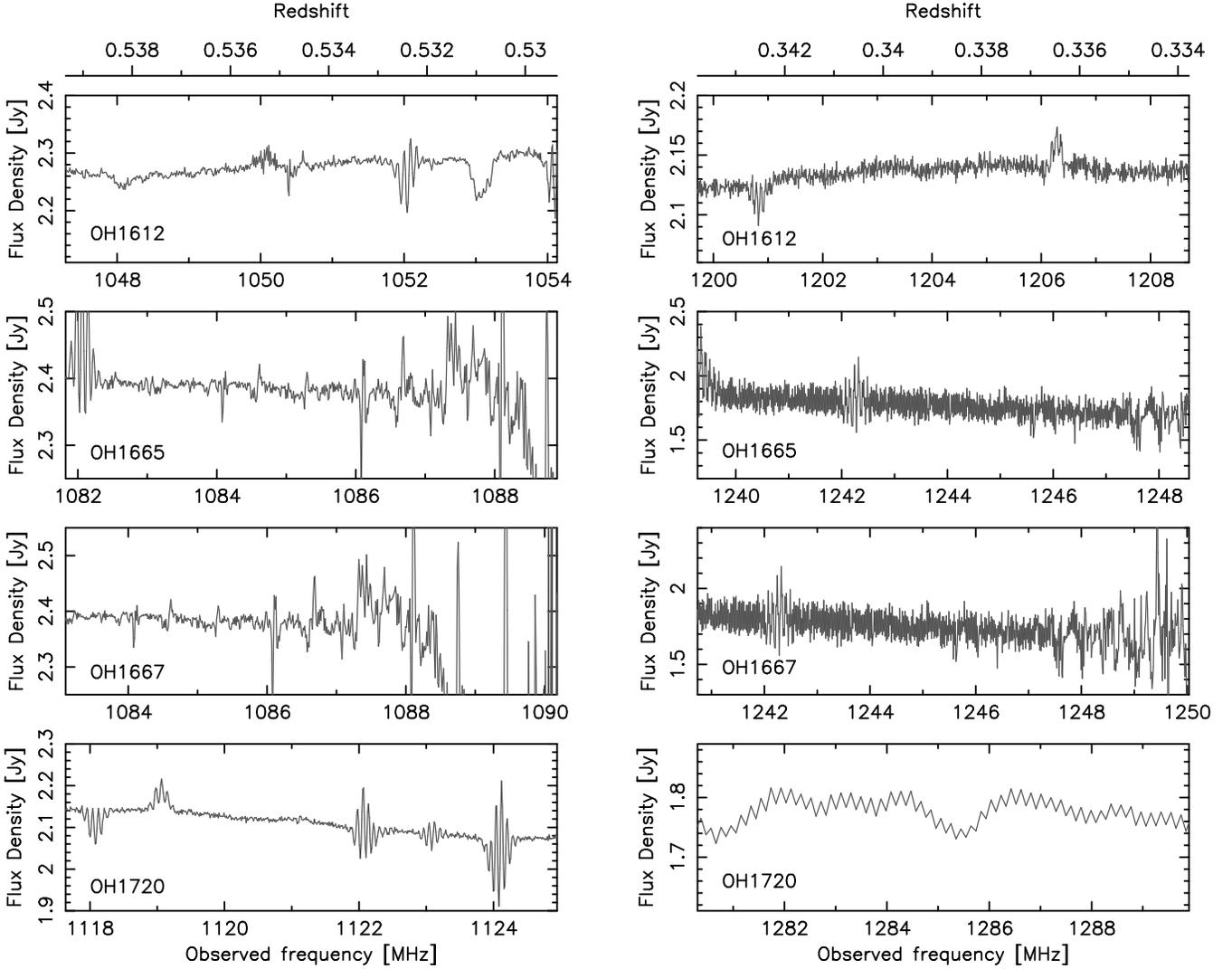}
\caption{The OH  data at frequencies close to the redshift of the \HI\ absorption, $z_{_{\rm HI}}\pm0.005$. 
The left panels show the expected frequencies of OH absorption for the $z=0.534$ absorption
feature and the right panels for the putative $z=0.339$ feature. All but the last panel are part of a 200 MHz
wide scan and so shown at the observed spectral resolution of 3--4 \kms. For 1720 MHz at $z=0.339$, the spectrum
is from a 800 MHz wide scan and shown at the observed resolution of 21 \kms.}
\label{oh-spectra}
\end{figure*}
From \HI\ 21-cm  at $z=0.5344$, we expect the 1665 and 1667 MHz  OH main lines at 1085.3985 and 1086.67 MHz
and the 1612 and 1720 MHz satellite lines at 1050.75 and 1121.33 MHz, respectively. Likewise, if real,
for  \HI\ 21-cm  at $z = 0.3388$, we expect the OH 18-cm lines at 1243.94 \& 1245.40 MHz (main) and 1204.22 \& 1285.11 MHz
(satellite). There is, however, no evidence of OH absorption in either case (Fig. \ref{oh-spectra}).

In Table \ref{abs} we list the best OH column density limits for these systems, together with those of the previous
searches. Based on the optical--near-infrared colour of $V-K = 10.26$, we expect molecular fractions of close to unity
\citep{cwc+11} or $N_{\rm OH}/N_{\rm HI}\gapp10^{-4}$ \citep{cwm+06}. This is ruled out for all of the known 21-cm
absorbers, except perhaps that associated with the host which has the weakest limit. Even so, given that only 46\%
of the band is not completely ruined by RFI (Tanna et al., in prep.), the presence of significant columns of molecular gas
elsewhere along this sight-line cannot be ruled out.
 
\subsubsection{Reddening by dust associated with atomic gas}

 Given that there is no evidence of molecular absorption coincident with the redshifts of the 21-cm absorbers, we revisit
the possibility that the reddening is due to dust associated with the intervening atomic gas \citep{cmr+98}. From all of
the published associated $z\geq0.1$ absorption searches, \citet{cw10} found a $3.63\sigma$ correlation between the
21-cm line strength and the $V-K$ colour. Unlike the molecular line strength correlation \citep{cwm+06}, this exhibits
considerate scatter, although it does comprise many more data points and is quite fragile, with the correlation
quickly disappearing as various sub-samples are removed.

\begin{figure}
\centering \includegraphics[angle=270,scale=0.77]{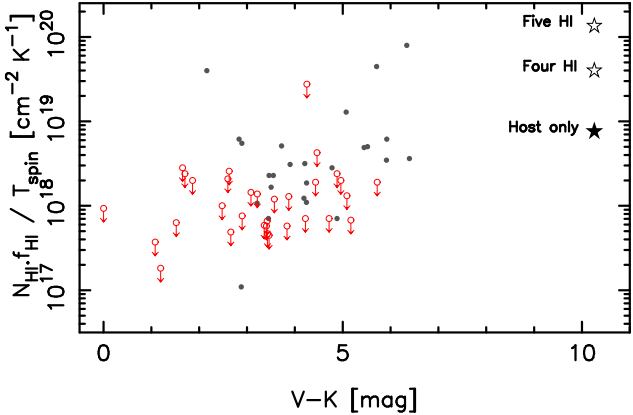}
\caption{The 21-cm line strength versus optical--near-IR colour for the associated $z\geq0.1$ absorbers for which the colours are available \citep{cw10}.
The filled circles are the 21-cm detections and the unfilled circles the non-detections.
The filled star shows the line strength for the $N_{\rm HI} = 7.5\times10^{18}.\,(T_{\rm spin}/f_{\rm HI})$ absorber associated with the host of  J0414+0534 and
the hollow stars the total of the line strengths for the four robust detections, and including the putative $z=0.339$ feature (Table \ref{abs}).}
\label{red3}
\end{figure}

Another point of note was that the 21-cm line strength of the associated ($z = 2.64$) absorber was weaker than expected,
based upon the trend defined by the other points (Fig. \ref{red3}).
Replacing this with the total column density so far observed along this sight-line (the unfilled stars in the figure)\footnote{Setting 
the values to $N_{\rm HI} = 2.8$ and $6.7\times10^{19}.\,(T_{\rm spin}/f_{\rm HI})$, in order to maintain consistency with the other measurements.}, moves the line strength
closer to the expected trend and, as before,  including the limits via the {\sc asurv} survival analysis package \citep{ifn86},
the total column density increases the
significance of the correlation to  $3.66\sigma$ for four absorbers and $3.70\sigma$ including the putative fifth
absorber. 

 This may be further evidence of dust associated with the dense atomic gas in the intervening systems being the cause of
the extremely red colour of this object, although the individual contributions of these absorbers toward the total
reddening of the quasar light cannot, as yet, be established. Note also, that most of the other sources have only been
searched for 21-cm close to the host redshift (i.e. for associated absorption). Therefore, strictly speaking, these too
represent lower limits, since further absorption systems along their sight-lines have not been ruled out. Furthermore,
as for the molecular gas, substantial atomic absorption may exist in this sight-line at redshifts corresponding to RFI
affected frequencies. In either case, the large number of 21-cm absorbers detected over the useful fraction of the band,
in conjunction with the apparent obscuration of the source, may have far reaching implications for the number of
gas-rich galaxies missed by optical surveys.

\section{Summary}

We have now completed the analysis of a full-band decimetre wave spectral scan towards the very red ($V-K = 10.26$)
quasar MG\,J0414+0534. From the $46$\% of the band not completely ruined by RFI, we have found three (possibly four)
strong intervening absorbers, in addition to the previously detected absorption in the host galaxy at $z=2.64$
\citep{mcm98}. At a total of four (or five), this represents a new record in the number of 21-cm absorbers found along a
single sight-line, the previous being a total of two systems towards PKS\,1830--211 \citep{lrj+96,cdn99}, which has  $V-K = 6.25$
(see \citealt{cwm+06}).

The new detections occur at redshifts of $z = 0.3388$ and $z=0.5344$, each being very strong absorbers with column densities of
$N_{\rm HI}\geq6.7\times10^{19}.\,T_{\rm spin}$ and $\geq2.8\times10^{19}.\,T_{\rm spin}$ \scm, respectively.
This qualifies both as damped Lyman-$\alpha$ absorption systems for a paltry $T_{\rm spin} \sim10$~K (an order of
magnitude lower than the lowest yet found, \citealt{ctm+07}). However, given the RFI in the spectrum at 1061 MHz, the
 $z = 0.3387$ feature requires confirmation.

Despite the large column densities, no OH absorption was found in either the main or satellite 18-cm lines at these redshifts,
although the very red colour of the background source suggests large molecular fractions somewhere
along the sight-line \citep{cwm+06}. Summing the observed \HI\ column densities does strengthen the atomic gas
abundance/$V-K$ colour correlation \citep{cw10}, although this remains fairly scattered. Therefore, dense molecular gas at an
RFI affected redshift and/or the possibility that J0414+0534 is intrinsically red cannot be ruled out.

However, given that this extremely red sight-line has yielded at least three intervening gas-rich galaxies (which would
likely have remained undiscovered through optical spectroscopy), does have implications for obscured galaxy populations.  With the
large field-of-view and instantaneous bandwidths that will be available with the Square Kilometre Array and the 
Australian SKA Pathfinder (ASKAP)\footnote{Which will scan \HI\ 21-cm absorption over the $z\lapp1$ redshift range discussed in Sect. \ref{sec:inc},
through the {\em First Large Absorption Survey in \HI} (FLASH).}, large-scale blind surveys of radio sources with faint optical counterparts will soon be possible, allowing
us to quantify the number of such galaxies hidden to optical surveys.
 
The Centre for All-sky Astrophysics is an Australian Research Council Centre of Excellence, funded by grant  CE110001020.


\end{document}